# Large-Scale Collective Entity Matching

Vibhor Rastogi
Yahoo! Research
rvibhor@yahoo-inc.com

Nilesh Dalvi
Yahoo! Research
ndalvi@yahoo-inc.com

Minos Garofalakis*
Technical University of Crete
minos@acm.org

## ABSTRACT

There have been several recent advancements in Machine Learning community on the Entity Matching (EM) problem. However, their lack of scalability has prevented them from being applied in practical settings on large real-life datasets. Towards this end, we propose a principled framework to scale any generic EM algorithm. Our technique consists of running multiple instances of the EM algorithm on small *neighborhoods* of the data and *passing messages* across neighborhoods to construct a global solution. We prove formal properties of our framework and experimentally demonstrate the effectiveness of our approach in scaling EM algorithms.

## 1. INTRODUCTION

*Entity Matching (EM)* is the problem of determining if two entities in a data set refer to the same real-world object. Entity matching is a complex and ubiquitous problem that appears in numerous application domains (including image processing, information extraction and integration, and natural language processing), often under different terminology (e.g., coreference resolution, record linkage, and deduplication). As an example, consider the following instance of the problem - this will be our running example throughout the paper.

EXAMPLE 1. *Consider a collection of paper publications obtained from multiple bibliography databases. Our goal is to determine which papers and authors from the different databases are the same entities. Each paper $p$ has a subset of the attributes $p.title$, $p.journal$, $p.year$, $p.category$, while each author $a$ has attributes $a.fname$, $a.lname$. We also have a set of relations defined over these entities:* $\mathcal{R} = \{Authored, Cites, Coauthor\}$, *where* $Authored(a, p)$ *denotes that* $a$ *is an author of paper* $p$, $Cites(p_1, p_2)$ *states that paper* $p_1$ *cites* $p_2$, *and* $Coauthor(a_1, a_2)$ *denotes that authors* $a_1$ *and* $a_2$ *have co-authored some paper. (Note that the Coauthor relation can actually be easily derived through a self-join on Authored.)*

*Work partially supported by the European Commission under FP7-PEOPLE-2009-RG-249217 (HeisenData).

Based on the above example, we formalize the entity matching problem as follows.

ENTITY MATCHING (EM) PROBLEM : Let $E$ denote a collection of *entities* (e.g. set of *papers* and *authors*). Each element $e$ in $E$ is characterized by a set of attributes, $e.A_1, e.A_2, \cdots, e.A_k$ (e.g. *title* and *journal* for *paper*, and *name* for *author*). Furthermore, we have a set of relations, $\mathcal{R} = R_1, \cdots, R_m$ over $E$ (e.g. *Authored* and *Cites*). The objective is to determine, given $e_i, e_j \in E$, whether $e_i$ and $e_j$ are the same entity in real world.

Conventional approaches to EM (dating back to the seminal work of Newcombe [15] and Fellegi and Sunter [8]) focused on discovering independent pair-wise matches of entities using a variety of attribute-similarity measures. These approaches ignored the relational information $\mathcal{R}$. Recently, several *collective entity matching* techniques have been developed that use the relational information like *Cites* and *Coauthor* to make all the matching decisions collectively. They have been shown to significantly outperform the conventional approaches in terms of accuracy. State-of-the-art Collective EM methods typically rely on sophisticated Machine Learning (ML) tools, e.g. Conditional Random Fields (CRFs) [14, 16], relational Bayesian networks [17], latent Dirichlet models [4, 9], and Markov Logic Networks [18]. We urge the reader to refer to Appendix D for a comprehensive classification of EM algorithms that have been considered in the literature along with their accuracy/efficiency trade-offs, and merits of collective algorithms.

Unfortunately, the collective EM techniques suffer from a fundamental problem: their poor scalability. The high cost of probabilistic inference over large EM graphs renders these methods computationally infeasible for large data sets.

**Our Contributions.** In this paper, we propose a principled framework for scaling state-of-the-art, Collective EM algorithms to large data sets. Our framework allows for general EM algorithms to be modeled as "blackboxes" that take in a set of entities along with a collection of *evidence*, and output a set of matches. Our framework approximates the run of the matcher on the entire data set by: (1) running multiple instances of the matcher on several small subsets of the entities, and (2) *Message-Passing*, i.e., passing a judiciously-built message-set across the instances to exchange information between different runs of the matcher. We present a formal analysis of our framework, demonstrating that, for a broad class of "well-behaved" entity matchers (satisfying natural "monotonicity" properties), the approach is *provably sound*. We also consider the case of Collective EM based on probabilistic models, and propose novel message-passing schemes that significantly improve EM recall without compromising soundness. The salient properties of our EM framework can thus be summarized as follows.

1. *Generic:* Any (deterministic or probabilistic) EM algorithm can be incorporated in our framework.

2. *Accurate & Provably Sound:* Our formal analysis show soundness for a broad class of EM techniques, while novel message-passing schemes guarantee high recall.

3. *Scalable & Parallelizable:* By running EM instances on small subsets of entities together with intelligent message passing, our framework is highly scalable and also suitable for a parallel implementation.

Our experimental study over real data sets verifies the effectiveness of our approach, demonstrating the ability to scale a state-of-the-art Collective EM algorithm based on Markov Logic Network [18] to real-life data sets of more than 58,000 references involving around 1.3 million matching decisions with little or no impact on the accuracy of the results.

# 2. OVERVIEW

In this section, we give an overview of our framework for scaling collective entity matchers. We illustrate our approach using the *Markov Logic Network (MLN)*-based entity matcher [18], which is a state-of-the-art collective EM technique. Section 2.1 gives an intuitive introduction to MLN matchers, describing how such a matcher works on a simple example. In Section 2.2, we use this MLN matcher to explain informally our scaling framework. Then, in Sections 3–5, we give the formal descriptions of the various components of our framework in more detail.

## 2.1 The MLN Entity Matcher

Consider a set of entities given in Figure 1, where each entity represents an author reference, and an edge between two entities represent a *Coauthor* relationship. Thus, we have $Coauthor(a_1, b_2)$, $Coauthor(a_2, b_3)$ etc. Further, using a string-similarity function, we define a relation called *Similar* on the entities. In this example, we have $Similar(a_1, a_2)$, $Similar(b_i, b_j)$ and $Similar(c_i, c_j)$ for $i, j = 1, 2, 3$.

We want to use the relations *Similar* and *Coauthor* to derive a relation *Match*, which tells us the entities that should be matched. Intuitively, we want to match two entities if they are similar and they have a matching coauthor. We can express this as the following two rules:

$$R_1 \; : \; Similar(x, y) \Rightarrow Match(x, y)$$
$$R_2 \; : \; Similar(x_1, y_1) \land Coauthor(x_1, x_2) \land Coauthor(y_1, y_2)$$
$$\land \; Match(x_2, y_2) \Rightarrow Match(x_1, y_1)$$

Note that these rules are *not hard constraints*, but rules which we believe are likely to hold. In the MLN matcher, one can write arbitrary rules and the system assigns a weight (a number) to each one of them. (These MLN rule weights are typically learned by using (labeled) training data.) The weights are used to assign a score to each possible set of entity matches: the score of a set is given by the *total weight* of all the rules in that set that become true. Although we will not go into the technical details of MLN, it is important to note that these MLN rule weights have a natural probabilistic interpretation, and the score assigned to a set of matches can be normalized to obtain a valid probability measure over possible match decisions [18]. Thus, the system tries to find the *highest-scoring set of entity matches* (which is also the set of matches with the highest probability).

As an example, let us assume that the system assigns a weight of -5 to $R_1$ and a weight of 8 to $R_2$. A negative weight for $R_1$ implies that the *Similar* relation by itself is not sufficient to match the two entities. On the other hand, since $-5 + 8 = 3$ is positive, entities that are similar and share a coauthor are matched. To see this, consider the entities $c_1$, $c_2$ for which $Similar(c_1, c_2)$ holds; Consider two possible sets of matches: (i) the set $\{Match(c_1, c_2)\}$, and (ii) the empty set $\{\}$. For the first set, $R_1$ is true (by putting $x = c_1$ and $y = c_2$) and $R_2$ is also true (by putting $x_1 = c_1, y_1 = c_2$, and $x_2 = y_2 = d_1$)[1]. Thus, the score assigned to this set is $-5 + 8 = 3$. On the other hand, the second (empty) set makes $R_1$ and $R_2$ both false, and is assigned a score of 0. So, the first set is assigned a larger score, and the system declares $c_1$ and $c_2$ to be matches.

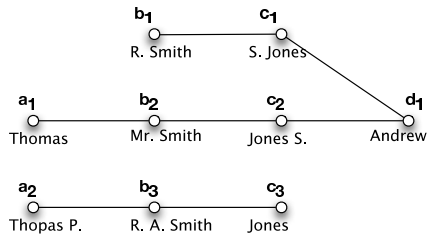

**Figure 1: An Instance of Entity Matching Problem: Nodes are author references and edges are coauthor relationships.**

In the above deduction, just by looking at $c_1$, $c_2$, and $d_1$, we decided that assigning $Match(c_1, c_2)$ leads to optimal solution. But, how do other entities in the graph affect this decision? One nice property of our MLN-based matcher is that $Match(c_1, c_2)$ is still the right decision in the globally-optimal solution, even as more entities are added to the graph. We call this the *monotonicity* property (formally defined in Section 3). It turns out that monotonicity is a general property that real-life entity matchers often exhibit, and our EM framework is designed to exploit this.

Continuing with our example in Figure 1, having matched $c_1$, $c_2$, the matcher also infers $Match(b_1, b_2)$ for similar reasons, since the match further increases the total weight by $8 - 5 = 3$. At this point, we cannot further increase the total weight of the solution by adding any new single match. However, the pairs $(a_1, a_2)$, $(b_1, b_2)$ and $(c_1, c_2)$ show an interesting property, which truly illustrates the collective nature of MLN. Setting any one of them as a match just adds a weight of -5; but, setting all three of them *collectively* increases the total weight, as $R_1$ fires three times (once on each pair) leading to a weight of -15 but $R_2$ also fires two times contributing 16. The net result is an increase in total weight by +1. Thus, our MLN matcher decides to match all three pairs. In general, MLN-based matchers consider all matching decisions collectively, and there could be large collections of entity pairs, where *matching all the pairs* in the set leads to an increase in the total score, but the advantage can only be seen when looking at the set of pairs as a whole (as in the example above).

## 2.2 Our Framework

We now give an overview of our scalable EM framework. It consists of three key components: (i) modeling an entity matcher as a *black box* (ii) running multiple instances of the matcher on small subsets of the entities (termed *neighborhoods*), and (iii) using *message passing* across the instances to control the interaction between different runs of the matcher. Below, we present these three key components informally, and also outline some key theoretical properties.

---
[1] Our MLN also specifies the trivial *reflexivity* rule (i.e., each entity is equal to itself), so that $Match(d_1, d_1)$ holds.

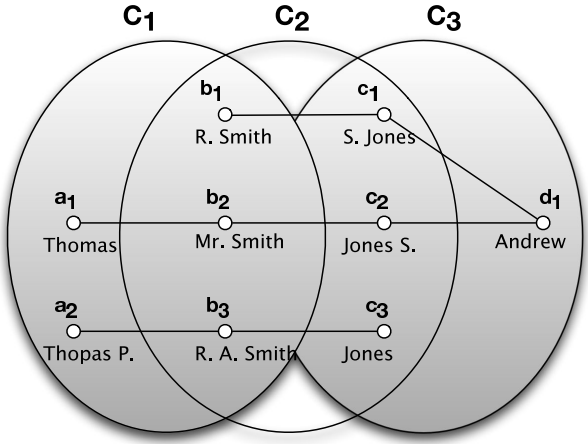

**Figure 2: A Cover of Entities.**

**Black-box EM Abstraction.** We consider a very generic abstraction for an entity matcher that takes a set of entities and outputs a set of matches. We define two kinds of matchers : a *deterministic matcher* that simply outputs a set of matches, and a *probabilistic matcher* that also outputs a set of matches, but uses a complete probability distribution over the set of matches to do so. (Formal definitions are in Section 3.) We also list a set of desirable properties of an entity matcher, which we can exploit in our framework. We call a matcher *well-behaved* if it satisfies these properties.

The deterministic abstraction is very general and can model any entity matcher. The probabilistic abstraction is a special case of a deterministic matcher, where the final output of the probabilistic matcher is the set of matches that has the highest probability. The reason for defining the probabilistic abstraction as a special case is that we can apply additional techniques when we know that the matcher is backed by a probability distribution. All known, state-of-the-art purely-collective matchers (which are the main focus of our scaling effort), are indeed probabilistic.

**Neighborhoods.** A *neighborhood* is simply a subset of the entities $E$ and a *cover* is a set of (potentially overlapping) neighborhoods whose union is $E$. An example cover of entities in Figure 1 is given in Figure 2: it describes a cover consisting of three neighborhoods, $C_1$, $C_2$, and $C_3$.

There is a vast amount of literature on *blocking* techniques [13], whose objective is to efficiently group together entities that are similar. The notion of cover extends the notion of blocking to group entities based on not just their similarity, but also on their relational closeness. For instance, in neighborhood $C_3$ shown in Figure 2, entities $d_1$ and $c_1$ are in the same neighborhood not because they are similar, but because they share a coauthorship edge. More generally, a neighborhood may even have entities of different types, e.g. an *"Author"* and a *"Paper"*. Section 4 describes covers formally.

**Message Passing.** We now come to the most important component of the framework : *message passing*. We start by analyzing how the MLN matcher described in Section 2.1 runs when applied separately to each of the three neighborhoods in Figure 2. In neighborhood $C_1$, MLN will not output any match. This is because none of the pairs $(a_1, a_2)$ and $(b_1, b_2)$ have enough evidence to get matched. If we try to match both pairs, we will incur a weight of +8 by rule $R_2$ but a weight of -10 by rule $R_1$ (which fires twice). So, it will decide not to match either of them. Similarly, it will not produce any match in $C_2$. In $C_3$, it matches the pair $(c_1, c_2)$ as shown in Section 2.1. Thus, the total set of matches produced by all the neighborhoods is simply $\{(c_1, c_2)\}$.

We notice that MLN misses several matches when run separately on each neighborhood. To recover these misses, we pass *messages* between neighborhoods. A *simple message* is just a set of matches found by a neighborhood. For instance, after $C_3$ runs, it passes a message to the neighborhood $C_2$ stating $Match(c_1, c_2)$. When $C_2$ runs MLN again with the new evidence $Match(c_1, c_2)$ added to its input, it is now able to match $Match(b_1, b_2)$. Thus, we recover one of the missed messages. The resulting scheme is called *simple message passing scheme* and is formally described in Section 5.

However, the simple message passing scheme cannot recover matches $(a_1, a_2)$, $(b_2, b_3)$ and $(c_2, c_3)$. This is because, as we showed earlier, matching the three pairs increases the overall score only if *all three of them* are matched. Thus, neither $C_1$ nor $C_2$ alone can match any of the pairs. Furthermore, simple messages are no help as no matches are found in either neighborhood. To overcome this problem, we extend the notion of a simple message. Although $C_2$ cannot match pairs $(b_2, b_3)$ and $(c_2, c_3)$, it can deduce that matching any one of them also results in the other pair being matched; thus, either *both pairs* should be matched or *none of the two pairs* should be matched. This is called a *maximal message*: It consists of a set of correlated matches, such that either all of them are true or none of them are. Intuitively, it represents a *"partial inference"* by a neighborhood, waiting to be completed. $C_2$ thus generates a maximal message $\{(b_2, b_3), (c_2, c_3)\}$. Similarly, $C_1$ generates a maximal message $\{(a_1, a_2), (b_2, b_3)\}$. These two messages, when combined, essentially *"complete the chain"*, and result in all three pairs being matched. The resulting scheme is termed *maximal message passing scheme*, denoted by MMP, and is developed in detail in Section 5.2.

### 2.2.1 Theoretical Properties

We now describe some key properties of the framework that we want to analyze, theoretically and empirically. Let $\mathcal{E}$ denote an entity matcher, $\mathcal{E}(E)$ denote the set of matches produced by $\mathcal{E}$ on entities $E$. Let $M \in \{\text{SMP}, \text{MMP}\}$ be a message passing technique and let $M(E)$ denote the set of matches produced by our framework using $M$. Note that the absolute precision/recall numbers of $M$ are not the right metric to evaluate our framework, since these numbers are tied to the underlying entity matcher $\mathcal{E}$. Instead, we define the following four properties.

1. *Soundness* : this is the fraction of matches in $M(E)$ that are also in $\mathcal{E}(E)$. We say that $M$ is *sound* if it has soundness 1, i.e., $M(E) \subseteq \mathcal{E}(E)$.

2. *Completeness* : this is the fraction of matches in $\mathcal{E}(E)$ that are also in $M(E)$. We say that $M$ is *complete* if it has completeness 1, i.e., $\mathcal{E}(E) \subseteq M(E)$.

3. *Consistency* : we say that $M$ is consistent if the final set of matches do not depend on the order in which neighborhoods are evaluated and messages are exchanged.

4. *Scalability* : we want the time complexity of $M$ to be low, preferably linear in the number of neighborhoods. Here, we assume that the sizes of neighborhoods are bounded, and study scalability with respect to the number of neighborhoods.

In Section 3, we define certain natural, intuitive properties on entity matchers and say that a matcher is *well-behaved* if it satisfies these properties. Our main result is as follows.

THEOREM 1. *If $\mathcal{E}$ is a well-behaved entity matcher, then both* SMP *and* MMP *are sound, consistent, and have time complexity linear in the number of neighborhoods.*

While there are no theoretical guarantees on completeness, we demonstrate empirically that our framework is in fact complete on the data sets we consider. In the following sections, we formally define entity matchers, the notion of *well-behaved* matcher, various message passing schemes, and revisit Theorem 1.

## 3. BLACK-BOX ABSTRACTION

We define two blackbox abstractions for an entity matcher. The first, which we call a *Type-I matcher*, is a deterministic matcher that outputs a set of matches. This is the most general abstraction that models any entity matching algorithm. The second abstraction, which we call a *Type-II matcher*, gives a probability distribution on sets of possible matches. As we shall see, a Type-II matcher is a special kind of Type-I. However, we introduce the second abstraction because we can apply additional techniques in our framework when the matcher is probabilistic. We describe below the two abstractions.

### 3.1 Type-I Matcher

Recall from Section 1 that an instance of entity matching problem consists of a set of entities $E$ along with a set of attributes for each entity and a set of relationships $\mathcal{R}$ defined on them. In the rest of the paper, we will use $E$ to denote a set of entities and implicitly assume the presence of attributes and relationships on entities.

DEFINITION 1 (TYPE-I MATCHER). *A Type-I matcher is a function* $\mathcal{E} : E \times 2^{(E \times E)} \times 2^{(E \times E)} \mapsto 2^{(E \times E)}$.

The first argument to $\mathcal{E}$ is a set of entities, the next two arguments are sets $V_+ \subseteq (E \times E)$ and $V_- \subseteq (E \times E)$, which are called the evidence sets. The set $V_+$ is a set of entity pairs that are known to be matches, while the set $V_-$ is a set of entity pairs that are known to be non-matches. The output of the function $O \subseteq E \times E$ is the set of pairs of entities declared as matches by the algorithm.

This definition treats the entity matching module as a complete black box except for a mechanism to provide evidence, which we need in our framework. We can model any entity matcher as a Type-I matcher: the sets $V_+$ and $V_-$ are simply ignored if not used by the entity matcher. We use $\mathcal{E}(E)$ to denote the output of the matcher when there is no evidence, i.e. when $V_+ = V_- = \emptyset$. Also, we use $\mathcal{E}(E, V_+)$ to denote the output when $V_- = \emptyset$.

We assume certain properties constraining how a matcher can use its evidence sets. First, we assume idempotence: the output of a matcher should not change if the output is itself again given as the positive evidence. We call this property idempotence.

DEFINITION 2 (IDEMPOTENCE). *A Type-I matcher is idempotent if for all* $E, V_+$ *and* $V_-$, *denoting* $O = \mathcal{E}(E, V_+, V_-)$, *the following holds:* $\mathcal{E}(E, O, V_-) = O$

Obviously a matcher that ignores its evidence sets, satisfies idempotence. In fact, we expect every matcher to satisfy idempotence. In addition to idempotence, we assume another monotonicity property: if we add more entities to the input entity set $E$ or give more positive evidence $V_+$, we get more matches in the output, while if we give more negative evidence $V_-$, we get less matches in the output.

DEFINITION 3 (MONOTONICITY). *A Type-I entity matcher is monotone if for any inputs* $E, V_+, V_-$ *and inputs* $E', V_+', V_-'$ *such that* $E' \supseteq E$, $V_+' \supseteq V_+$, *and* $V_-' \supseteq V_-$ *the following hold: (i)* $\mathcal{E}(E', V_+, V_-) \supseteq \mathcal{E}(E, V_+, V_-)$, *(ii)* $\mathcal{E}(E, V_+', V_-) \supseteq \mathcal{E}(E, V_+, V_-)$, *and (iii)* $\mathcal{E}(E, V_+, V_-') \subseteq \mathcal{E}(E, V_+, V_-)$.

We illustrate using examples in the Appendix A that many entity matching techniques in the literature satisfy monotonicity. At the same time, it should be noted that even non-monotone matchers can be scaled in our framework albeit without the theoretical soundness guarantee.

DEFINITION 4. *We say that a matcher is* well-behaved *if it satisfies idempotence and monotonicity.*

### 3.2 Type-II Matcher

DEFINITION 5 (TYPE-II MATCHER). *A Type-II matcher $\mathcal{E}$ is a function that takes as input a set of entities $E$ and associates a probability distribution $P_\mathcal{E}$ on the set $2^{E \times E}$.*

The probability of a set $S \subseteq E \times E$, $P_\mathcal{E}(S)$, is the probability that exactly the pairs in $S$ are matches.

Note that we have not specified an explicit mechanism to specify the initial evidence sets. This is because, since the matcher gives a complete probability distribution, any evidence sets $V_+, V_-$ can easily be incorporated by conditioning the distribution $P_\mathcal{E}$ on matches in $V_+$ to be true and $V_-$ to be false.

The output of a Type-II matcher $\mathcal{E}$ is the set $S \in 2^{E \times E}$ that has the highest probability according to the distribution $P_\mathcal{E}$. We denote it by $\mathcal{E}(E)$. There can be more than one most likely set, in which case, we will prefer a set with the largest size. More precisely, $\mathcal{E}(E)$ is any set $S \subseteq 2^{E \times E}$ that satisfies the following two properties : (i) for all $S'$, $P_\mathcal{E}(S) \geq P_\mathcal{E}(S')$ and (ii) for all $S'$ such that $P_\mathcal{E}(S) = P_\mathcal{E}(S')$, we have $|S| \geq |S'|$.

In presence of evidence sets $V_+$ and $V_-$, we can incorporate it in the entity matcher by simply conditioning the probability distribution $P_\mathcal{E}$ on $V_+$ and $V_-$. Thus, we define $\mathcal{E}(E, V_+, V_-)$ to denote any set among the largest most likely sets that contains all matches in $V_+$, but none from $V_-$.

PROPOSITION 1. *A Type-II matcher is also a Type-I matcher, i.e. it satisfies the idempotence axiom.*

We also define a supermodularity property on Type-II matchers that is related to the monotonicity property of Type-I matchers.

DEFINITION 6 (SUPERMODULARITY). *An entity matcher $\mathcal{E}$ is supermodular if for all entity pair sets $S, T \subseteq E \times E$ such that $S \subseteq T$, and for all entity pairs $p \in E \times E$, we have*

$$\frac{P_\mathcal{E}(T \cup \{p\})}{P_\mathcal{E}(T)} \geq \frac{P_\mathcal{E}(S \cup \{p\})}{P_\mathcal{E}(S)}$$

Intuitively, supermodularity says that the matches are positively correlated: increasing the set of matches from $S$ to to a larger set $T$ only increases the likelihood of another match $p$. Thus supermodularity is essentially just the monotonicity requirement stated for probabilistic matchers.

PROPOSITION 2. *Let $\mathcal{E}$ be a supermodular Type-II matcher. Then, $\mathcal{E}$ is a monotone Type-I entity matcher.*

Supermodularity is a well-studied property of probability distributions (see [11]). In fact, exact inference techniques are known for Markov Networks when the clique potentials are supermodular [11]. However, these work only if the cliques are edges (while we support any general entity matcher), and have super-linear complexity (quadratic or worse). Thus such techniques are not well-suited for extremely large data sets that we are interested in.

## 4. COVERING

In this section, we define the notion of a cover, and introduce a class of covers called *total covers*, which are especially suited for our framework. Given a set of entities $E$, a *cover* of $E$ is simply a set $\mathcal{C} = \{C_1, C_2, \cdots, C_k\}$ such that $E = C_1 \cup \cdots \cup C_k$. Each $C_i$ is called a *neighborhood*.

For example, consider the set of entities shown in Figure 1 with relations $\mathcal{R} = \{Coauthor, Similar\}$ over $E$ as defined in Section 2.1. Let $C_1$, $C_2$ and $C_3$ be the sets as shown in Figure 2. Then, $\mathcal{C}_1 = \{C_1, C_2, C_3\}$ is a cover. The set $\mathcal{C}_2 = \{C_1, C_3\}$ is also a cover.

Next we define the notion of total cover that extends the notion of a cover. Given a relation $R$ over $E$, and a subset $C \subseteq E$, let $R(C)$ denote the subset of the relation induced by $C$, i.e. the set of tuples in $R$ that solely consists of entities from $C$. Thus, $R(E)$ is the complete relation.

**DEFINITION 7** (TOTAL COVER). *A cover $\mathcal{C} = C_1, \cdots, C_k$ is a total cover of a set of entities $E$ w.r.t a set of relations $\mathcal{R}$ if for all relations $R \in \mathcal{R}$, we have $R(E) = R(C_1) \cup \cdots \cup R(C_k)$*

For example, in Figure 2, $\mathcal{C}_2 = \{C_1, C_3\}$ is not a total cover, since the tuple $Coauthor(b_1, c_1)$ is not contained in either of the two neighborhoods. However, one can verify that the cover $\mathcal{C}_1 = \{C_1, C_2, C_3\}$ is a total cover.

Total cover is a desirable property in our framework, since tuples in R that are not included in any neighborhood get *"lost"*, as they do not participate at all in the matching process. Given any cover, we can make it a total cover as follows: Define the boundary of a neighborhood $C$ to be the set of entities $e$ for which there is an entity $e'$ in $C$ such that both $e$ and $e'$ occur together in some tuple in $\mathcal{R}$. Then, by expanding each neighborhood to also include its boundary, we obtain a total cover. Total covering is a natural extension of the notion of *blocking*, used extensively in entity matching [13]. Blocking divides entities into neighborhoods such that for any pair of similar entities, there is a neighborhood that contains the pair. Thus, blocking is a total covering over the *Similar* relation. When we have other relations, like *Coauthor* and *Cites*, we can have neighborhoods containing entities that are not similar, and even entities of different types, e.g an author and a paper.

In this work, we construct a total cover by first constructing a total cover over *Similar* relation using the Canopies algorithm [13], and then taking the boundary of each neighborhood with respect to other relations. Because of the boundary, a neighborhood will have entities of multiple types and entities which are dissimilar. E.g. the use of *Coauthor* edges brings together authors whose names are very dissimilar to the same neighborhood.

## 5. MESSAGE PASSING

Now we describe the message passing algorithm, which is at the heart of our framework. The input is a cover $\mathcal{C} = C_1, C_2, \cdots C_k$ of entities $E$ and an entity matcher $\mathcal{E}$. We describe two message passing schemes: SMP, that works for any matcher, and MMP, that is more advanced but only works with probabilistic matchers.

### 5.1 Simple Message Passing Scheme (SMP)

We described the algorithm informally in Section 2.2. The algorithm maintains a set $A$ of active neighborhoods, which are the neighborhoods that can potentially produce new matches. $A$ is initialized to the set of all neighborhoods. As the matcher $\mathcal{E}$ is run on a neighborhood $C$, the neighborhood is removed from $A$. For the run, the matches found so far are given as evidence. If the run finds new matches, all the neighborhoods that are affected by the new

---

**Algorithm 1** SMP **Inputs:** Entity Matcher $\mathcal{E}$, A cover $\{C_1, \ldots, C_s\}$

1: $A \leftarrow \{C_1, \ldots, C_s\}$ // set of active neighborhoods
2: $M_+ \leftarrow \emptyset$ // set of matches found
3: **while** $A \neq \emptyset$ **do**
4:     $C \leftarrow$ remove a neighborhood from $A$
5:     $M_C = \mathcal{E}(C, M_+)$
6:     $A = A \cup Neighbor(M_C - M_+)$
7:     $M_+ = M_+ \cup M_C$
8: **end while**
9: return $M_+$ as the set of final match.

---

matches are put back into the set of active neighborhoods. The algorithm terminates when $A$ becomes empty. The pseudo-code for the algorithm is given in Algorithm 1.

**THEOREM 2.** *Let $\mathcal{E}$ be any well-behaved Type-I matcher. Then,*

1. *(Convergence)* SMP *terminates after finite steps.*
2. *(Soundness) The set of matches produced $\subseteq \mathcal{E}(E)$.*
3. *(Consistency) The output of* SMP *is invariant under the order in which neighborhoods are evaluated.*

**Running Time** Let $n$ be the number of neighborhoods, $k$ be the maximum size of each neighborhood, and let $f(k)$ be the running time of $\mathcal{E}$ on a neighborhood of size $k$. For a neighborhood $C$, the maximum number of matches it can produce is $C \times C$, which has cardinality at most $k^2$. Each time $C$ is added to the set $A$, the set $(C \times C) - M_+$ strictly decreases. Thus $C$ can be added to $A$ at most $k^2$ times. Hence, we get the following result:

**THEOREM 3.** *The time complexity of* SMP *is $O(k^2 f(k)n)$, where $n$ is the number of neighborhoods, $k$ is the maximum size of a neighborhood and $f(k)$ is the running time of $\mathcal{E}$.*

Assuming that the size of neighborhoods is bounded, the time complexity is linear in the number of neighborhoods (and hence, linear in the number of entities). In practice, a neighborhood is never evaluated $k^2$ times, and the running time is much lower than the theoretical upper bound.

### 5.2 Maximal Message Passing Scheme (MMP)

Now we describe the maximal message passing scheme (MMP) for Type-II probabilistic matchers.

**Motivation** Consider the pairs $(a_1, a_2)$, $(b_2, b_3)$ and $(c_2, c_3)$ in Figure 2. We showed in Section 2.1 that matching all the three pairs increases the overall score if and only if all the three of them are matched. Thus applying SMP leads to a 'chicken and egg' problem : $(i)$ $MLN$ can match $(a_1, a_2)$ and $(b_2, b_3)$ when run on $C_1$ if and only if it is given $(c_2, c_3)$ as evidence, and $(ii)$ $MLN$ can match $(b_2, b_3)$ and $c_2, c_3)$ when run on $C_2$ if and only if it is given the match $(a_1, a_2)$. For SMP to discover the three matches, we need both $(i)$ to happen before $(ii)$, and $(ii)$ to happen before $(i)$. Consequently, none of the three matches are output by the SMP algorithm. The maximal message passing scheme described below overcomes this problem.

**Maximal messages** A maximal message is a set of pairs such that either all of them will be matched by the matcher $\mathcal{E}$ or none of them will be matched. Thus the set $\{(a_1, a_2), (b_2, b_3), (c_2, c_3)\}$ in the above example is a maximal message.

**DEFINITION 8** (MAXIMAL MESSAGE). *A maximal message w.r.t matcher $\mathcal{E}$ is a set of pairs $M \subseteq E \times E$ such that $M \subseteq \mathcal{E}(E)$ or $M \cap \mathcal{E}(E) = \emptyset$.*

The following proposition follows from the definition.

PROPOSITION 3. (i) *If $M$ is a maximal message then so is any $M' \subseteq M$. (ii) If $M$ and $M'$ are two maximal messages s.t. $M \cap M' \neq \emptyset$, then $M \cup M'$ is also a maximal message.*

---

**Algorithm 2** COMPUTEMAXIMAL (**Inputs:** Entity Matcher $\mathcal{E}$, Neighborhood $C$, Evidence $M_+$)

1: For each pair $p = (e_1, e_2)$ of entities in $C$, compute $\mathcal{E}(C, M_+ \cup \{p\})$
2: Construct a graph $G(V, E)$ defined as follows:
   - $V$ contains a node corresponding to each pair $p$ of entity references in $C$.
   - $E$ contains an edge between pairs $p$ and $p'$ if $p \in \mathcal{E}(C, M_+ \cup \{p'\})$ and $p' \in \mathcal{E}(C, M_+ \cup \{p\})$.
3: For each connected component of $G$, output a message consisting of all the pairs in the component.

---

**Computing maximal messages** Next we describe an algorithm that computes a set of maximal messages for a neighborhood $C$. Intuitively, for each pair of matches, using a call to $\mathcal{E}$ with the pair as evidence, all other pairs entailed by it are computed. Groups of pairs connected by this process constitute the maximal messages. The pseudo-code is given in Algorithm 2.

**Passing maximal messages** The MMP algorithm passes maximal messages between neighborhoods. The idea is that two maximal messages can combine to produce a set of sound (i.e. correct) matches, if they together increase the probability of the solution. In our example, neighborhood $C_2$ produces a maximal message $\{(b_2, b_3), (c_2, c_3)\}$ and neighborhood $C_1$ produces a maximal message $\{(a_1, a_2), (b_2, b_3)\}$. When these two messages are combined, the new maximal message has enough evidence to match all its three pairs.

---

**Algorithm 3** MMP **Inputs:** Entity Matcher $\mathcal{E}$, cover $\mathcal{C} = \{C_1, \ldots, C_k\}$

1: $A \leftarrow \{C_1, \ldots, C_s\}$ // set of active neighborhoods
2: $M_+ \leftarrow \emptyset$ // set of matches found
   $T \leftarrow \emptyset$ // set of maximal messages found
3: **while** $A \neq \emptyset$ **do**
4:   $C \leftarrow$ remove a neighborhood from $A$
5:   $M_C = \mathcal{E}(C, M_+)$
     $T_C = ComputeMaximal(C, M_+)$
6:   $M_+ = M_+ \cup M_C$
     $T = (T \cup T_C)^*$
7:   **Find sound maximal messages:** Check if there exists $M \in T$ s.t. $P_\mathcal{E}(M_+ \cup M) \geq P_\mathcal{E}(M_+)$. If yes, then update $M_+ = M_+ \cup M$.
8:   $A = A \cup Neighbor(M_C \cup M - M_+)$
9: **end while**
10: return $M_+$ as the set of final match.

---

We now formally define the MMP algorithm. The pseudo-code for the algorithm is given in Algorithm 3. The algorithm, similar to SMP, maintains a set of active neighborhoods. It maintains a set $M_+$ of matches found so far. In addition, it maintains a set $T$ of maximal messages found so far. When a neighborhood is processed (step 5), we find both the new matches $M_C$ and the new maximal messages $T_C$ (using Algorithm 2). Then (step 6), we use Proposition 3 to combine maximal messages from two neighborhoods. Proposition 3 says that a set of maximal messages can be equivalently represented by replacing overlapping messages with their unions. Given a set of (possibly overlapping) maximal messages

$T$, let $T^*$ denote the set obtained from $T$ by repeatedly picking two messages from $T$ that overlap and replacing them with their union, until all the messages are disjoint. Thus, we update the set $T$ by taking its union with the new maximal messages $T_C$ and then taking $(T \cup T_C)^*$. Finally (step 7), we check if a maximal message can be converted to a sound message.

Step 7 is the only step that requires $\mathcal{E}$ to be a probabilistic matcher, as it involves a check $P_\mathcal{E}(M_+ \cup M) \geq P_\mathcal{E}(M_+)$.

Algorithm 3 assumes that we can efficiently compute $P_\mathcal{E}$ for any set of inputs. This condition, in general, is true for probabilistic models, where finding the optimal solution, i.e. $\arg\max_S P_\mathcal{E}(S)$ is a very expensive operation that involves searching over a large space, but computing $P_\mathcal{E}(S)$ for a specific $S$ is very cheap using the parameters of the model.

**Theoretical Guarantees** We revisit Theorem 1 for MMP.

THEOREM 4. *Let $\mathcal{E}$ be a supermodular Type-II matcher. Then, MMP exhibits convergence, soundness, and consistency.*

**Running Time** Let $n$ denote the number of neighborhoods, $k$ be the maximum size of each neighborhood, and let $f(k)$ be the running time of $\mathcal{E}$ on a neighborhood of size $k$. As in the case of SMP, each neighborhood can be processed at most $k^2$ times. However, unlike SMP, processing a neighborhood is more expensive for MMP. It involves two things: a call to COMPUTEMAXIMAL and a computation of $(T \cup T_C)^*$. We prove:

THEOREM 5. *MMP runs in time $O(k^4 f(k)n)$, where $n$ is the number of neighborhoods, $k$ is the maximum size of a neighborhood and $f(k)$ is the running time of $\mathcal{E}$.*

Similar to SMP, the time complexity is linear in the number of neighborhoods. Again, in practice, a neighborhood is never evaluated $k^2$ times, and the running time is much lower than the theoretical upper bound.

# 6. EXPERIMENTS

We evaluate the performance, both accuracy and efficiency, of our message passing schemes. For evaluation, we use the collective entity matching algorithm of Singla et al. [18], which is a state of the art algorithm, and uses Markov Logic Networks. We call this matcher MLN. Additionally, in Appendix C, we evaluate a second matcher based on the declarative framework for entity matching [2] that uses soft collective rules written in a datalog like language. We call that RULES. RULES is a Type-I matcher, since its not probabilistic, while MLN is a Type-II probabilistic matcher. Appendix B contains the exact set of rules, entity types and relations used by each matcher. Appendix A analyses the properties of various matchers. In particular, the matchers we use are monotonic and supermodular.

For covers, we use the *Canopy* cover given by McCallum et al. [13]. Our message passing algorithms are written in Java. All our experiments are run on an Intel Xeon® 2.13GHz machine running linux with 4GB RAM.

**Datasets** We use two datasets. The HEPTH dataset is a dataset used in the 2003 KDD Cup and available for download from the KDD website. It contains information on papers in theoretical high-energy physics. It has 58,515 author references mentioned across 29,555 unique papers and 13,092 unique authors. It contains completely labeled ground truth.

Since no other large labeled dataset is publically available, we manually prepared a second dataset, DBLP, as follows. From the

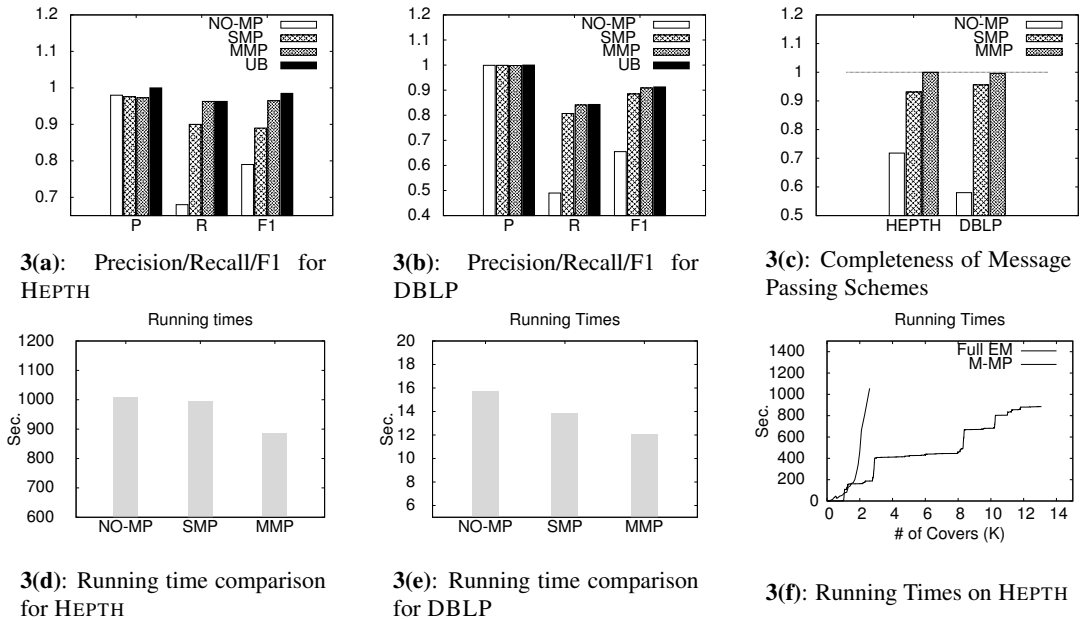

**3(a):** Precision/Recall/F1 for HEPTH

**3(b):** Precision/Recall/F1 for DBLP

**3(c):** Completeness of Message Passing Schemes

**3(d):** Running time comparison for HEPTH

**3(e):** Running time comparison for DBLP

**3(f):** Running Times on HEPTH

**Figure 3: Results for** MLN **matcher**

DBLP Computer Science Bibliography [12], we obtained publications corresponding to a large number of Database and AI conferences. We compiled 19,408 papers with a total of 50,195 author references, with 21,278 distinct authors. However, since DBLP data is clean, we manually add noise by randomly adding small mutations to author names. Thus, the original names serve as the ground truth in evaluating this dataset.

## 6.1    Accuracy of message passing schemes

**Precision/Recall for** MLN **Matcher.** In the HEPTH dataset, the covering algorithm results in 13K neighborhoods containing a total of 1.3M entity pairs. We compare three algorithm: NO-MP is the no message passing scheme, where MLN is run on each neighborhood but no messages are exchanged. SMP is the simple message passing scheme. MMP is the maximal message passing scheme. For each scheme, we compute the precision, recall and the $F1$ score, which is the harmonic mean of their precision and recall.

We also want to compare these numbers with those for running MLN on the whole dataset. However, running MLN at this scale is infeasible. For instance, on HEPTH, it involves creating a Markov Logic Network with 1.3M nodes and running inference over it. Instead, we use the following technique to obtain an upper bound on the set of matches that MLN can produce. For each entity pair, we give the MLN algorithm the ground truth about all other entity pairs and run the matcher to decide the given entity pair. Since our matcher satisfies the supermodularity property, we can show that this is indeed an upper bound on the set of matches that MLN can produce. We call this scheme UB. Note that this is not an algorithm since it uses the ground truth.

Recall of UB is a provable upper bound on the recall of full run of MLN, because of the supermodularity property. However, the precision of UB might be lower. So, to compute an upper bound on $F1$, we take the recall value of UB and a maximum precision 1.

Fig. 3(a) plots these scores for various schemes on HEPTH. In each of the message passing schemes, the precision is very high and close to 1. This is because all the message passing schemes are sound, and the original MLN algorithm itself has a high precision. The precision of the MLN is however not 1, and this explains why MMP can have lower precision than SMP, and even lower precision than NO-MP. As we approach the true output of MLN, the precision also approaches the true precision of MLN.

The $F1$ of MMP comes close to UB. Note that the UB need not be attainable by any message passing scheme, as the original EM may have a recall less than that of UB and precision less than 1.

Figure 3(b) shows the same graph for DBLP. The DBLP dataset produces 30K neighborhoods containing a total of 0.5M entity pairs. Note that while both the datasets have roughly the same number of author references, DBLP produces twice the number of neighborhoods as HEPTH, with a much smaller average neighborhood size (as evident by the less number of entity pairs). This is due to the fact that DBLP always contains the full author names as opposed to HEPTH, where names are often abbreviated, leading to more name clashes and fewer neighborhoods of larger size.

**Soundness/Completeness for** MLN **Matcher.** We know that our message passing schemes are sound, but not necessarily complete. Hence, we study the completeness of various algorithms empirically. Note that completeness is different from recall: recall is measured with respect to the ground truth, and is an intrinsic property of the underlying matcher. Completeness is the fraction of the matches found by a particular message passing scheme compared to running EM on the entire dataset holistically, and is a property of our framework. Again, since we do not have the true set of matches given by MLN on the whole dataset, we use the matches of UB to obtain a lower bound on completeness. Figure 3(c) shows the completeness for various schemes with respect to UB. We see that MMP has completeness 1 for HEPTH. Thus, MMP is both sound and complete and hence, gives the exact same output as running the EM on the whole HEPTH dataset. Similarly, for DBLP, MMP has completeness nearly 1.

## 6.2    Running Times

Figure 3(d) shows the running times of the various message pass-ing schemes on HEPTH using MLN matcher. We observe an interesting and counter-intuitive behavior. SMP, which does the extra work of passing messages as well as running neighborhoods multiple times, has a lower running time than NO-MP, that does not do message passing at all. MMP, that passes more messages, has an even lower running time. This apparent paradox can be explained as follows. In all the three schemes, the total running time is dominated by the sum of running times of MLN on all the neighborhoods. The actual overhead of message passing is minimal. However, SMP wins over NO-MP because, since neighborhoods share entities, messages often reduce the *active size* of the neighborhoods. All the pairs in a neighborhood that have already been matched by other neighborhoods are not part of inference anymore. Hence, MLN runs faster. For instance, if a neighborhood has 100 entity pairs to consider, and each pair has already been matched by some previous neighborhood, EM does not need to do any work on this neighborhood. Since MLN has a non-linear complexity, multiple small neighborhoods can be processed much faster that a single big neighborhood. The net result is that even though SMP has to revisit neighborhoods multiple times, the *active sizes* of neighborhoods keep getting lower, and the total running time is lower. MMP, by a more effective message passing, achieves an even lower running time by the same principle. Thus a better message passing scheme not only increases the precision/recall of the matching, it also results in a lower running time.

Figure 3(e) shows the running times for MLN on DBLP dataset. DBLP exhibits a similar behavior. However, we observe an interesting difference. While both the datasets have roughly the same size, the running times on DBLP is an order of magnitude lower. This is because the neighborhoods in DBLP are much smaller, owing to reasons explained earlier. As a result, MLN runs significantly faster on the neighborhoods in DBLP.

We also analyze the running time as a function of the size of the input. Figure 3(f) shows, for each $k$, the total running time of MLN when run on the first $k$ neighborhoods together. The graph shows an exponential behavior of MLN, and it becomes prohibitively expensive to run MLN on more that 2500 neighborhoods. On the other hand, our message passing scheme, MMP, exhibits a linear behavior for the whole 13,000 neighborhoods. At certain points, we see jumps in the curve of MMP. This is because some of the neighborhoods are large, and whenever a new large neighborhood is included, the running time shows a small jump.

## 6.3 Parallelizing EM

We now demonstrate how to parallelize our EM framework. The idea is to run it in rounds. All neighborhoods are marked active at the beginning. In each round, EM is run on all the active neighborhoods in parallel, then the new evidence from the runs is collected, and used to obtain active neighborhoods for the next round. E.g., suppose we have 10000 neighborhoods, and 10 available machines. We randomly assign 1000 neighborhoods to each machine, and run EM. Suppose at the end of first round 500 neighborhoods are active. We redistribute them again among the 10 machine, so that each runs 50 neighborhoods in parallel.

We implemented this parallel framework on *Hadoop*, which is an open-source *Map-Reduce* based grid framework. Each round of our algorithm is run using one *Map* and one *Reduce* job as follows. The *Map* job assigns each neighborhood to a grid machine where EM is run on it. The output is used by the *Reduce* job to bring all the new evidence for each neighborhood together. Each new *Map* job redistributes the active neighborhoods to grid machines.

To evaluate the parallel implementation, we constructed a third dataset much bigger that the previous two datasets. This dataset consists of the entire collection of publications from DBLP bibliography data, which we call DBLP-BIG. The dataset has 4,606,712 author references among 2,303,254 publications. DBLP-BIG produces 1,723,190 neighborhoods, with a total of 41,713,259 entity-pairs that need to be resolved. We used the grid implementation of our framework to run MLN on DBLP-BIG over a Hadoop installation in Yahoo! with 30 machines. We also ran DBLP-BIG over a single machine to compare the speedup due to parallelism. The following table summarizes the running time of our algorithms.

|  | NO-MP | SMP | MMP |
|---|---|---|---|
| Single Machine | 208 | 329 | 285 |
| Grid (30 machines) | 18 | 30 | 27 |

**Table 1: Running Times on Grid (minutes)**

On all the three variants of message passing, the speedup is around 11. There are couple of reasons we do not see a perfect speedup of 30. First, the grid has some overhead in setting up the mapper and reducer jobs on all the nodes. The second reason is purely statistical. Since neighborhoods are "randomly" assigned to nodes, there is a statistical skew in the assignment, and some nodes get multiple bigger than average neighborhood. There is ongoing research in the community on reducing skew in *MapReduce* that can further improve our speedup. Nevertheless, we have demonstrated that our techniques can run efficiency on grid and achieve good speedup.

# APPENDIX

## A. MONOTONICITY OF MATCHERS

Many of the entity matching algorithms proposed in the literature satisfy our monotonicity properties. Recall that monotonicity says that the output of the matcher changes monotonically if new entities are added to the entity set $E$ or new evidence is added to the evidence sets $V_+$ and $V_-$. However, note that we do not require monotonicity to hold when new relationships are added between existing entities.

The non-probabilistic matchers in [5] and [6] use an iterative approach in which newly found matches can only lead to more matches. These matchers satisfy the monotonicity properties.

Further, the state-of-the-art Markov Logic based matcher [18], referred to as MLN(B) in [18], satisfies the monotonicity and supermodularity properties. The proof for MLN(B) follows from the following general result.

> **PROPOSITION 4.** *If the rules in a Markov Logic Network have only one* Match *term in the implicant, then the resulting matcher satisfies the monotonicity and super-modularity properties.*

For example the rule $R_2$ (See 2.1) has only one *Match* term on the implicant side and hence satisfies monotonicity. Other variants of the MLN matcher were also proposed in [18] by using different rules. We note that all the rules in [18] except for the transitivity rule (i.e. $equals(A, B) \wedge equals(B, C) \Rightarrow equals(A, C)$) satisfy the condition above, and hence all variants that do not include this rule are monotonic. The transitivity rule is not monotonic and our framework does not provide formal soundness guarantee for this rule. However, if transitive closure is applied at the end of matching, it maintains the monotonicity property of the matcher. In other words, the transitive closure of any monotonic matcher is monotonic. Thus, transitive closure can be supported by taking a simple transitive closure at the end of each iteration of message passing.

Several works have looked at using rules to collectively match entities. In a recent work [2], a declarative framework called *Dedupalog* was proposed, where datalog rules are specified by the user which act as constraints for entity matching. Rules can be hard, such as the first rule below that matches two authors $x$ and $y$ if they are known to be equal based on an externally specified predicate $AuthorEQ$. Rules can also be soft, such as the second rule below, which says that if two entities do not have a matching coauthor, then they are unlikely to be the same.

$$equals(x, y) \quad \Leftarrow \quad AuthorEQ(x, y)$$
$$\neg equals(x, y) \quad \Leftarrow \quad \neg(Wrote(x, P_1), Wrote(y, P_2),$$
$$Wrote(x', P_1), Wrote(y', P_2),$$
$$equals(x', y'))$$

Given a set of rules, the goal of the dedupalog matching algorithm is to instantiate the *Match* predicate such that (i) no hard rule is violated, (ii) the number of violated soft rules is minimized, and (iii) *Match* is transitively close, i.e. $equals(x, y) \wedge equals(y, z) \Rightarrow equals(x, z)$.

Again, the transitive closure property as a constraint may violate the monotonicity of the entity matcher. However, we have the following proposition.

> **PROPOSITION 5.** *Let* Dedupalog* *be the fragment of dedupalog without negation and transitivity constraint. Then,* Dedupalog* *is monotone.*

Note that negation does not always lead to non-monotonicity. E.g., if we have a rule $\neg a \Leftarrow \neg b$ containing negations, we can rewrite it to an equivalent rule $b \Leftarrow a$ that does not have negations. In fact, all the rules discussed in [2] are monotone.

## B. DETAILS OF MLN AND RULES

In this section, we mention the rules used in the experimental evaluation for both MLN and RULES. Both the matchers have three predicates: (i) $similar(e_1, e_2, score)$, which gives the similarity $score$ between a pair of authors, (ii) $coauthor(e_1, e_2)$, which contains pairs of entity references that have coauthored a paper, and (ii) $equals(e_1, e_2)$ which is the predicate that we want to compute. The similarity scores between two authors was computed using the *JaroWrinkler* distance, and was discretized to the set $\{1, 2, 3\}$ with 3 being the highest possible similarity.

**MLN** For the MLN matcher, we used the Alchemy [18] system to learn the weights of the rules using training data. Below we give the rules along with the weights that were learnt.

```
1   similar(e_1, e_2, 1) ⇒ equals(e_1, e_2)        -2.28
2   similar(e_1, e_2, 2) ⇒ equals(e_1, e_2)        -3.84
3   similar(e_1, e_2, 3) ⇒ equals(e_1, e_2)        12.75
4   coauthor(e_1, c_1) ∧ coauthor(e_2, c_2)         2.46
    ∧ equals(c_1, c_2) ⇒ equals(e_1, e_2)
```

The first three rules describe the effect of string similarity on the match decision if we ignored co-author relationships. A similarity score of 1 and 2 is not enough to match the pair $p$, as evident by the negative weights associated with the first two rules, while a score of 3 means that the names are very similar and we get a positive weight leading to a match. The final rule explains the effect of co-authors relationship. It gives a positive weight for matching a pair of authors if they have a pair of matching coauthors. Section 2.1 explains how an MLN works using these type of rules.

**Rules** Unlike in MLN, where predicates have potentials and a set of predicates are selected if their joint potential is positive, we need to give explicit set of rules in RULES. We specify the following set of rules, inspired by the learnt MLN model.

1. $similar(e_1, e_2, 3) \Rightarrow equals(e_1, e_2)$
2. $similar(e_1, e_2, 2) \wedge coauthor(e_1, c_1) \wedge coauthor(e_2, c_2)$
   $\wedge\ equals(c_1, c_2) \Rightarrow equals(e_1, e_2)$
3. $similar(e_1, e_2, 1) \wedge coauthor(e_1, c_1) \wedge coauthor(e_2, c_2)$
   $equals(c_1, c_2) \wedge coauthor(e_1, c_3) \wedge coauthor(e_2, c_4)$
   $equals(c_3, c_3) \wedge \{c_1, c_2\} \neq \{c_3, c_4\} \Rightarrow equals(e_1, e_2)$

The first rule says that similarity 3 is a good evidence for matching. For similarity 2, we need at least one common coauthor and for similarity 1 we need at least two common coauthors. Finally, we use the 3-approxiamte algorithm in [2] to evaluate the above set of rules without transitive closure, followed by a transitive closure at the end.

## C. EXPERIMENTS ON RULES MATCHER

**Precision/Recall/Completeness** We now evaluate our framework with the RULES matcher on the two datasets. Recall that since RULES is a Type-I matcher, and does not generate probabilities, the MMP scheme is not applicable here. Hence, we evaluate the results of NO-MP and SMP. Although RULES is a collective matcher, it runs much faster than MLN, and we were easily able to run it on

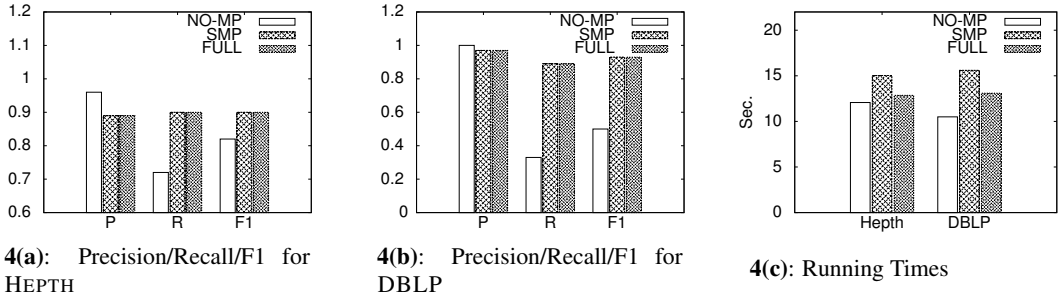

**4(a):** Precision/Recall/F1 for HEPTH

**4(b):** Precision/Recall/F1 for DBLP

**4(c):** Running Times

**Figure 4: Results for RULES matcher**

the entire datasets. Thus, we were able compute the soundness and completeness of the message passing schemes exactly. Figure 4(a) and (b) show the precision, recall, and the F1 measure for RULES on HEPTH and DBLP respectively. Here, FULL denotes running the matcher on the entire datasets holistically. On both the datasets, the precision and recall of SMP matched the full run, i.e. SMP was able to achieve completeness. The overall accuracy of RULES matcher is a bit lower than MLN.

**Running times** Figure 4(c) shows the running time of RULES on the two datasets. We would like to point out here that RULES is already a fast matcher and can be run directly on the whole data. Our main purpose of studying RULES was to demonstrate that SMP can easily handle rule-based matchers with soundness and completeness. However, even for a fast matcher like RULES, there is merit in running it with SMP, as it gives a natural way to parallelize any EM algorithm, and run it on grid on really huge datasets. In Appendix. 6.3, we have experiments showing the ability of our framework to run parallelly on a grid of computers. Implementing RULES directly on grid is non-obvious. Also note that RULES was designed with efficiency as the goal, and hence can only support a small fragment of datalog. Message passing opens up the possibility for supporting more complex rules by providing support for non-linear algorithms. Finally, unlike in the case of MLN, the running time of SMP is not lower than that of NO-MP, since RULES have linear complexity, and the savings due to running on neighborhoods with smaller active sizes does not compensate for the cost of revisiting neighborhoods.

## D. A BRIEF SURVEY OF ENTITY MATCHING TECHNIQUES

In this section, we give a brief survey of existing EM techniques and examine their accuracy/efficiency trade-offs. We use Example 1 to illustrate the principles behind various techniques.

• *Non-Relational Approaches:* Initial approaches to EM focused on pair-wise attribute similarities between entities. Newcombe [15] and Fellegi and Sunter [8], gave the problem a probabilistic foundation by posing EM as a classification problem (i.e., deciding a pair to be a match or a non-match) based on attribute-similarity scores. The bulk of follow up work on EM then focused on constructing good attribute-similarity measures (e.g., using approximate string-matching techniques) [3]. For instance, in Example 1, an attribute-based matcher decides to match two authors by estimating the similarity of their $fname$ and $lname$ strings. A fundamental shortcoming of attribute-based matchers is that they cannot perform disambiguation, e.g., if there are two different authors with the same name "J. Doe", they cannot be identified as separate entities.

• *Simple Relational Approaches:* Entity matching can be significantly improved by using relational information in addition to at-

tribute similarities. Simple relational techniques take the attributes of *related entities* into account when computing the similarity scores of entity pairs [1, 10]. For instance, in Example 1, when attempting to match two authors, looking at the attribute values of their co-authors and the set of authors that they cite can improve the prediction accuracy by a large margin. For example, two authors "J. Doe" and "John Doe" are more likely to be the same person if they have written papers with "M. Smith" and "Mark Smith", respectively. While still making matching decisions in a pair-wise manner, the use of relational information can help alleviate the problem of disambiguation. However, one potential pitfall is that "M. Smith" and "Mark Smith" might actually be two different authors with similar names.

• *Collective Approaches:* Collective approaches to EM make use of the fact that various matching decisions are interrelated. For instance, in the scenario described above, the match decision for "J. Doe" and "John Doe" is closely tied to the match decision for "M. Smith" and "Mark Smith". Thus, collective approaches do not just look at additional attribute values obtained through relational information; instead, they exploit *related match decisions* when trying to determine a given entity match.

Collective EM techniques can be further classified into two sub-lasses, *iterative* and *purely-collective* approaches. As their name suggests, iterative approaches [5, 6] iterate over the set of current matches, and, as match decisions are made, they are used to trigger further match decisions. For instance, if there is strong evidence to match "M. Smith" and "Mark Smith", this decision in turn can be used to match "J. Doe" and "John Doe" by exploiting the co-authorship relation. While intuitively simple, iterative approaches have the problem of *bootstrapping:* if we initially start from an empty set of matches, we might never have enough evidence to perform any match. On the other hand, purely-collective approaches avoid the bootstrapping problem by building sophisticated models of the interrelationships of related match decisions, and using these models to make match decisions in a truly-collective manner. For example, while neither of the ("M. Smith", "Mark Smith") and ("J. Doe", "John Doe") pairs might have strong evidence by itself to be declared a match, a purely-collective matcher may be able to declare both of them as matches collectively, due to their mutually-reinforcing relation (through co-authorship). Purely-collective approaches represent the current state-of-the-art in terms of matching quality (i.e., the accuracy of matches), and are typically based on recent ML advances in statistical relational learning. Such state-of-the-art EM tools rely on various various advanced probabilistic models, including Conditional Random Fields (CRFs) [14, 16], relational Bayesian networks [17], latent Dirichlet models [4, 9], and Markov Logic Networks [18]. Probabilistic models and inference provide a clean and principled way to model relational information and perform purely-collective EM; unfortunately, the high cost of probabilistic inference over large EM models has hitherto rendered such methods infeasible for large data sets.

**Scaling EM Algorithms** The issue of scalability arises even in the simple case of independent pair-wise entity matchers, due to the obviously quadratic complexity of the all-pairs comparisons. This problem has been addressed in the literature using techniques that rely on *blocking* [13]. Blocking methods try to group entities together based on simple, heuristic grouping criteria (e.g., by the initial letter of authors' last names), so that matching entities are very likely to fall under the same group. Instances of the EM algorithm are then run on each group individually and matching results are collected across all the groups. In general, blocking groups can overlap, and several different grouping criteria can be used (to increase the probability that matching entities are paired-up in some EM instance). For independent pair-wise entity matchers, blocking helps avoid the quadratic cost of all-pairs comparisons, essentially reducing the complexity of EM on $n$ entities from $O(n^2)$ to $O(kn)$, where $k \ll n$ denotes the average size of a group [13]. More recently, Whang et al. [7] have proposed an *iterative blocking* framework, where the key idea is to allow blocks to communicate their "local" EM results in an iterative manner (until a fixpoint is reached). As they demonstrate, sharing local matches allows for both better accuracy (by enabling some collective EM decisions across groups) and better runtimes (by avoiding duplicated matching effort); still, their development is, for the most part based on heuristics and does not provide any formal guarantees on the quality of the EM results. In addition, [7] does not consider the more complex case of *probabilistic* Collective EM, that, as discussed earlier, significantly exacerbates the scalability issues involved.

# E. PROOFS

## E.1 Simple Message Passing

We now give the proof for Theorem 2. We first restate it below

THEOREM 6 (THEOREM 2). *Let $\mathcal{E}$ be any well-behaved Type-I matcher. Then the following hold for* SMP*:*

1. *(Convergence)* SMP *terminates after finite steps.*
2. *(Soundness) The set of matches produced $\subseteq \mathcal{E}(E)$.*
3. *(Consistency) The output of* SMP *is invariant under the order in which covers are evaluated.*

**Proof:** Part (1) follows easily, since every time a cover is added to the set $A$ of active covers, the set $M$ strictly increases. Since the size of $M$ is bounded by $|E \times E|$, the algorithm must terminate.

To show (2), we need more results. We first define the notion of a sound message. Given a set $M \in E \times E$, we say that $M$ is *sound* if $M \subseteq \mathcal{E}(E)$, i.e. $M$ represents correct set of matches with respect to the matcher. Next we state the following result.

PROPOSITION 6. *Let $\mathcal{E}$ be a well-behaved matcher, $C$ be any neighborhood, and $M^+$ be a sound set of matches. Then $M_C = \mathcal{E}(C, M_+)$ is sound.*

**Proof of Prop. 6:** As $\mathcal{E}$ is monotone and $C \subseteq E$, we know that $\mathcal{E}(C, M_+) \subseteq \mathcal{E}(E, M_+)$. As $M_+$ is sound, we know that $M_+ \subseteq \mathcal{E}(E)$. Thus, using monotonicity, we get

$$\mathcal{E}(E, M_+) \subseteq \mathcal{E}(E, \mathcal{E}(E)) = \mathcal{E}(E)$$

where the last equality follows from the idempotence property. Combining, we get $M_C = \mathcal{E}(C, M_+) \subseteq \mathcal{E}(E, M_+) \subseteq \mathcal{E}(E)$ showing that $M_C$ is sound. □

Finally, using Proposition 6, we prove part (2) of Theorem 2. We use an induction argument. Assume that $M_+$ is sound at the beginning of some iteration (base case is true as $M_+ = \emptyset$ for the first iteration). The above Proposition guarantees that the $M_C = \mathcal{E}(C, M_+)$ is sound. Thus the new $M_+$, which is the union of $M_+$ and $M_C$ is again sound proving the induction step and completing the proof of part (2).

## E.2 Maximal Message Passing

We now give the proof for Theorem 4. We first restate it below

THEOREM 7 (THEOREM 4). *Let $\mathcal{E}$ be a supermodular Type-II matcher. Then,* MMP *always terminates. Moreover on termination, the output $M_+$ is sound (i.e. $M_+ \subseteq \mathcal{E}(E)$) and consistent (i.e. invariant under the order in which covers are run)*

As the MMP algorithm uses the COMPUTEMAXIMAL algorithm to compute the maximal messages, we first show the correctness of the COMPUTEMAXIMAL algorithm.

LEMMA 1. *Let $\mathcal{E}$ be any well-behaved type-I matcher, $M_+$ be any sound evidence set, and $C$ be any neighborhood. Then all the messages produced by* COMPUTEMAXIMAL *are maximal.*

**PROOF.** Let $M$ be any message in the output corresponding to a connected component $CC$ of the graph $G$. We shall show that $M$ is maximal. For this we use induction.

Base case: The set $M = \{p, p'\}$ is maximal if there is an edge between $p$ and $p'$ in $G$. To prove this recall the definition of maximality. For showing maximality of a two element set $M = \{p, p'\}$, w.l.o.g, it is sufficient to show that: $p \in \mathcal{E}(E, M_+)$ implies $p' \in \mathcal{E}(E, M_+)$. Assume the contradiction, i.e. $p \in \mathcal{E}(E, M_+)$ but $p' \notin \mathcal{E}(E, M_+)$. Since $p \in \mathcal{E}(E, M_+)$, the set $\{p\}$ is sound. Moreover $M_+$ is also sound implying that the set $M_+ \cup \{p\}$ is sound. Since $p$ and $p'$ are connected by an edge in $G$, we know that $p' \in \mathcal{E}(C, M_+ \cup \{p\})$. Applying Proposition 6(b) for the sound message $M_+ \cup \{p\}$ and neighborhood $C$, we get that $p'$ is also sound. This means that $p' \in \mathcal{E}(E, M_+)$, which contradicts our assumption that $p' \notin EN(E, M_+)$. Thus, the set $M_+ = \{p, p'\}$ is maximal.

Inductive hypothesis: Assume $M$ is maximal for all connected components $CC$ having at most $k$ nodes.

Inductive Step: Decompose any connected component $CC$ of $k+1$ nodes into two connected components of $CC_1$ and $CC_2$, such that $CC_1$ and $CC_2$ share a node and have at most $k$ nodes each. This is always possible for any connected component $CC$. By inductive hypothesis $M_{CC_1}$ and $M_{CC_2}$ are both maximal. Moreover since $CC_1$ and $CC_2$ share a node, the messages $M_{CC_1} = \cup_{p \in CC_1}\{p\}$ and $M_{CC_2} = \cup_{p \in CC_2}\{p\}$ have non-empty intersection. Applying Proposition 3, we obtain that $M = M_{CC_1} \cup M_{CC_2}$ is also maximal. Hence proved. □

Using Lemma 1, we give an informal proof of the Theorem 4.

**Proof of Theorem. 4** (Informal) The proof is similar to the inductive proof of soundness of for the SMP Algorithm. In MMP, there is just one extra step (Step 7) that affects $M_+$. In Step 7, we add the pairs in $M$ to the existing sound message $M_+$. All we need to show is that $M$ is sound. We ensure that the update done in Step 6 preserves the soundness of $M_+$, and hence the theorem will follow in a way similar to the proof for SMP.

To show $M$ is sound, note that $P_{\mathcal{E}}(M_+ \cup M) \geq P_{\mathcal{E}}(M_+)$. Let $O = \mathcal{E}(E)$. By supermodularity of $\mathcal{E}$, we know that $\frac{P_{\mathcal{E}}(O \cup M)}{P_{\mathcal{E}}(O)} \geq \frac{P_{\mathcal{E}}(M_+ \cup M)}{P_{\mathcal{E}}(M_+)} \geq 1$. Thus $P_{\mathcal{E}}(O \cup M) \geq P_{\mathcal{E}}(O)$, which is only possible if $M \subseteq O$. Hence $M$ is sound. This completes the proof. □